\documentstyle[preprint,aps,epsf]{revtex}
\global\firstfigfalse
\global\firsttabfalse
\begin{document}
\bibliographystyle {prsty}
\title{
Zero--Bias Anomaly in Finite Size Systems
}
\author{Alex Kamenev and Yuval Gefen }
\address{
Department of Condensed Matter Physics, The Weizmann Institute of Science,
Rehovot 76100, Israel. 
}

\maketitle

\begin{abstract}
The small energy anomaly  in the single particle density of states of 
disordered interacting systems is studied for the zero dimensional case. 
This anomaly interpolates between the non--perturbative 
Coulomb blockade and the perturbative limit, the latter being an extension of 
the Altshuler--Aronov zero bias anomaly at $d=0$. 
Coupling of the  zero dimensional system to a
dissipative environment leads to effective screening of the interaction and a 
modification of the density of states.

\end{abstract}


\section{Introduction}
\label{s1}

The density of states (DOS) of disordered {\em interacting} electronic systems 
exhibits singularity near the Fermi energy. 
This phenomenon manifests itself in the suppression of the tunneling conductance 
at small external bias and is referred to as zero bias anomaly (ZBA). 
It has been explained  and investigated theoretically by  
Altshuler and Aronov (AA) in 1979 for $d=3$ and latter 
extended for lower dimensionality by Altshuler  Aronov and Lee 
\cite{Altshuler79}(for a review see 
Ref.~\cite{Altshuler85}). They have predicted that for a d--dimensional system 
with diffusive disorder and weak short range interactions 
the singular correction to the DOS behaves as  
$\delta\nu_d\sim (\mbox{max}\{\epsilon,T\})^{d/2-1}$ with the dimensionality 
$d=1,2,3$. 
For three dimensional ($d=3$) samples the ZBA constitutes a small but nonanalytic 
correction to the DOS (as long as the dimensionless conductance, $g$, 
is large, $g\gg 1$). 
It  may  be argued subsequently that this phenomenon is a precursor of the ``Coulomb 
gap'', which appears in the Anderson insulator regime, $g<1$.  
Throughout this paper we shall consider good conductors, hence  
we shall not discuss here the localized regime. But remarkably enough, even 
for good 
metals, $g\gg 1$, there is a situation where the ZBA is a strong effect. 
This is the case with a finite size system. The original treatment of AA was 
given for infinite (quasi) d--dimensional systems. 
For a finite size system the quantization of the spectrum of the diffusive 
modes  becomes of crucial importance. For instance, the electron--electron 
interaction term in the Hamiltonian takes the form 
\begin{equation}
H_{int}=\frac{1}{2}\sum_{\bf Q}V({\bf Q}):\rho_{\bf Q}\rho_{-{\bf Q}}:,
                                          \label{Hint}
\end{equation}                                           
where ${\bf Q}=2\pi{\bf n}/L$ are quantized momenta, $\rho_{\bf Q}$ is the 
${\bf Q}$ component of the electron  density operator and $:\ldots:$ denotes  
normal ordering. The quantization of 
the momentum has  two major consequences for the ZBA. 

First, if one accounts for all finite ${\bf Q}$, excluding the 
${\bf Q}=0$ mode, in the interaction 
Hamiltonian,  the singularity in the DOS is rounded off on the scale 
$\mbox{max}\{\epsilon,T\}\approx E_c$, where $E_c$ is the Thouless 
correlation energy. As long as  the ZBA is still 
a small correction to the total DOS on the scale $E_c$, 
$\delta\nu_d(E_c)\ll\nu^{[0]}$, the 
finite ${\bf Q}$ modes {\em do not} enhance the singularity for 
$\epsilon,T< E_c$. It is easy to see \cite{Altshuler85} that the requirement 
$\delta\nu_d(E_c)\ll\nu^{[0]}$ is satisfied for $g\gg 1$; in this case 
the finite ${\bf Q}$ contribution to the ZBA is small for {\em any} 
energy and temperature. It means that for such systems the finite ${\bf Q}$ 
interaction may indeed be  treated by perturbation theory, which 
produces regular expansion in powers of $1/g$. Throughout this paper we 
shall  assume that the condition $g\gg 1$ is satisfied, hence  
the perturbative treatment of the non--zero modes is applicable. 

The second consequence of the spectrum quantization is the special role 
played by the ${\bf Q}=0$ term in the interaction Hamiltonian. Its 
contribution to the  energy may be written as $V(0):N^2:/2$, where 
$N\equiv \rho_0$ is the total number of electrons in a dot. This term  in 
fact  corresponds to 
the classical charging energy of the dot. It leads to a  
strong singularity in the DOS. The  first order in interaction perturbative 
result is  $\delta\nu_0(\epsilon=0)= -V(0)/(4T)$ 
(for the ${\bf Q}=0$  contribution $T$ and $\epsilon$ 
are not interchangeable). In the  
limit $T< V(0)$, the perturbative treatment is not sufficient. We present 
here a  
relatively simple  method which allows us to treat ${\bf Q}=0$ contribution to the 
ZBA exactly. An exact solution is possible due to the fact that the  
zero--mode 
interaction term commutes with the total Hamiltonian. As a result 
the problem is trivial, although the zero--mode interaction has some 
interesting consequences. The thermodynamical and the 
response functions are practically unaffected by the 
zero--mode interactions (apart from modifying the  
statistical ensemble from grand canonical to canonical). 
The single particle DOS is modified in an essential way. We use 
imaginary time functional integral to integrate out all fermionic 
degrees of freedom. Similar methods were used in the context of Josephson 
junctions with  dissipative environment (for a review see \cite{Shon90}). 
Non--perturbative treatment reveals the exponential suppression of the 
DOS at the Fermi energy. This result reproduces the classical treatment 
of the ``orthodox'' theory of the Coulomb blockade \cite{Likharev89}. 
In other words, the $d=0$ generalization of the AA ZBA, and 
the Coulomb blockade, are two limiting cases of the same theory. 

This statement is in fact not new. In his original paper, 
Ref.~\cite{Nazarov89}, Nazarov had established the  connection 
between the two (including a  non--perturbative treatment of 
the finite ${\bf Q}$ modes). A similar, although more transparent, 
theory has recently been put forward by 
Levitov and Shytov \cite{Levitov95}. To a large extent, our results can be 
extracted as the $d=0$ limit of the expressions found in 
Refs.~\cite{Nazarov89,Levitov95}. The point is that 
Refs.~\cite{Nazarov89,Levitov95} used some uncontrolled (although plausible) 
approximations. We have shown that for ${\bf Q}=0$ all calculations may be 
done exactly, confirming some of the approximations made in Refs. 
\cite{Nazarov89,Levitov95}. 

One may combine the zero and finite ${\bf Q}$ mode interactions to the ZBA. 
In doing so we take into account the zero mode in an exact fashion and add on 
top of it the perturbative contribution of the ${\bf Q}\neq 0$ modes. As was 
mentioned above the latter can be expressed as a regular expansion in powers 
of $1/g$ valid at any temperature energy and dimensionality. To address the 
insulating regime one should consult Refs.~\cite{Nazarov89,Levitov95}. In the 
metallic limit the ${\bf Q}\neq 0$ contribution modifies the DOS at high 
energies ($\epsilon\geq e^2/2C$), as a precursor to the Coulomb blockade 
which dominates at lower energies (cf. Fig. \ref{f4}b).

The new ingredient in our analysis is the inclusion of  the 
zero--point (and thermal) motion of the electromagnetic environment in 
contact with the  
system. We show that the influence  of the environment may lead to a  
Debye screening of the zero--mode interaction, hence to a softening of the 
Coulomb blockade. The physics behind the zero--mode screening is the 
following: the total charge 
on the dot, $eN$, interacts with charges in the environment, leading to  
their redistribution. The polarized charge of the environment reduces 
the energy cost of adding (or removing)  an electron to (or from) the dot. 
There is a certain finite time constant (the $RC$ -- 
constant of the circuit), characterizing this redistribution of the 
environmental charge, rendering the effective interaction  in the dot 
non--instantaneous (retarded). It might appear that the  results obtained 
through this analysis are 
highly non--universal and depend on the particular choice of the model for 
the environment. We stress, however, that our result expressed by 
Eqs.~(\ref{GT}) and (\ref{Sscr}) is quite general, the only 
model--dependent feature is the concrete form of the screened zero--mode 
interaction, $V(\omega)$. Nonuniversality in this sense is unavoidable, 
due to the long range nature of the Coulomb interaction: the behavior of 
the dot depends on numerous long distance   features. 

A particularly interesting  case is when the   zero--mode 
interaction is fully screened in the long time limit. 
By this we refer to a situation when  on a long time scale the addition of an 
electron to the dot does not cost any energy (for instance there is a slow 
continuous leakage of any extra charge from the dot). 
In this case the usual Coulomb gap in the spectrum is absent and one obtains 
the power--law ZBA, very similar to the one known from  the physics of 
Luttinger liquid. 
The exponent is determined by the time scale of the environment 
polarization. This result  
for a quantum dot coincide with those obtained in 
Refs.~\cite{Devoret90,Glazman90} for a single tunnel junction connected 
to a  linear $RCL$ circuit \cite{foot1}. 
For a quantum dot one may consider the setup with 
only partially screened interaction, where the  gap in the DOS at low  
energies crosses over to a power--law ZBA at larger energies. 

Let us list several aspects of the problem, which we do not consider here. 
Although we deal with  finite size systems, we do not consider effects 
related to the discreteness of the single electron spectrum. It means that the 
mean level spacing, $\Delta$, is assumed to be the smallest energy scale in our 
problem. We restrict ourselves to the case of good metals, where  
$g=E_c/\Delta\gg 1$. Thus the energy interval of interest, 
$\Delta<\epsilon<E_c$, is wide. Since we are not interested in the 
single electron spectrum quantization, one may choose any boundary conditions 
for the electron wave functions. We prefer to use periodic boundary 
conditions  and employ  the momentum representation. 
We also stress that our analysis omits the underlaying periodicity of 
the problem as function of the charge of the positive background (or of a 
gate voltage) which could be manifested as a sequence of Coulomb blockade 
resonances. Instead we restrict ourselves to the case where the 
background is such that the dot prefers to have an integer charge 
(half way between resonances). 
One can easily generalize the same treatment for any background charge, 
noting that in  the close vicinity of a half--integer (near a resonance) a very 
different treatment, accounting for   multiple tunneling 
events \cite{Matveev91} is required. In fact even far from resonances 
multiple tunneling may 
be important (e.g ``inelastic co--tunneling'' \cite{Averin92}). 
For the sake of clarity  we restrict ourselves to a ``golden 
rule'' scenario, where we consider only lowest order processes in the 
tunneling amplitude, postponing the  
consideration of multiple tunneling to a further publication. 

The outline of the article is as follows. In Section \ref{s2} we recall the 
derivation of the ZBA given by AA and extend it to $d=0$. 
The non--perturbative treatment of zero-mode interaction is discussed in 
Section \ref{s3}, where we study in some details  the case of 
an instantaneous zero--mode interaction and rederive the results of 
the ``orthodox'' 
Coulomb blockade. Section \ref{s4} is devoted to the study of 
screened retarded 
zero--mode interaction. We show that in  one  particular case the ZBA 
reproduces the 
results derived for a single junction connected to a linear 
circuit.

\section{Zero--Bias Anomaly}
\label{s2}

\begin{figure}[htbp]
\epsfysize=4cm
\begin{center}
\leavevmode
\epsfbox{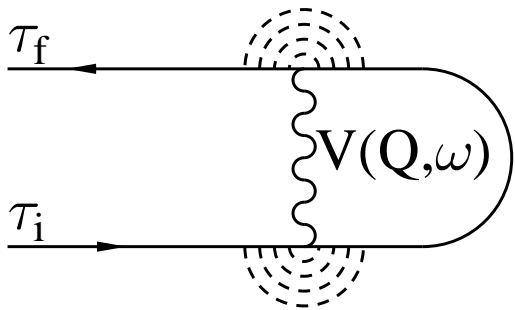}
\end{center}
\caption{\label{f1}  
First order interaction correction to the Green 
function; wavy line -- interaction; dashed lines -- impurity dressing 
(diffusons).}
\end{figure}

The derivation of the ZBA in the diffusive interacting systems proceeds as 
follows \cite{Altshuler85}. One calculates the {\em single particle} 
(not to be confused with the {\em thermodynamic}, $dn/d\mu$) 
density of states as 
function of energy or temperature. The single particle DOS is defined as 
the imaginary part of the trace of the single particle Green function
\begin{equation}
\nu(\epsilon)=-\left. \frac{1}{\pi}\Im\mbox{Tr}{\cal G}(\epsilon_n)
\right|_{i\epsilon_n\rightarrow\epsilon+i\delta}.
                                                 \label{dos}
\end{equation}                                                 
Following AA \cite{Altshuler85}, we   calculate the first order 
correction in the {\em screened} interaction to the single particle 
DOS in a dirty system. The dominant contribution \cite{foot2}  
comes from the exchange diagram 
depicted in Fig.\ \ref{f1}. 
After performing the fast momentum summation  one obtains 
\begin{equation}
\delta\nu(\epsilon)=-\frac{\nu^{[0]}}{\pi}\Im\,  T \!\!\!\!\!
\sum_{\omega_m>\epsilon_n} \sum_{\bf Q} \left. 
\frac{2\pi i V({\bf Q},\omega_m)}
{(D{\bf Q}^2+|\omega_m|+\gamma_{in})^2} \right|_
{i\epsilon_n\rightarrow\epsilon+i\delta} ,
                                                          \label{AA}
\end{equation}
where $D$ is the diffusion constant; $\gamma_{in}$ --  
the inelastic  relaxation rate; 
$V({\bf Q},\omega_m)$ -- the screened Coulomb interaction, given by 
\begin{equation}
\left[ V({\bf Q},\omega_m)\right]^{-1}=\left[ V^{[0]}({\bf Q})\right]^{-1}+
\Pi({\bf Q},\omega_m)
                                                          \label{scr}
\end{equation}
Here $V^{[0]}({\bf Q})$ is the bare (instantaneous)
Coulomb  interaction and $\Pi({\bf Q},\omega_m)$ is the polarization 
operator of the system.  
For an isolated diffusive system the polarization is 
\begin{equation}
\Pi({\bf Q},\omega_m)=
\nu^{[0]}\frac{D{\bf Q}^2}{D{\bf Q}^2+|\omega_m|} ; 
                                                          \label{pol}
\end{equation}
$\nu^{[0]}$ is  the DOS of non--interacting system. 

\begin{figure}[htbp]
\epsfysize=5cm
\begin{center}
\leavevmode
\epsfbox{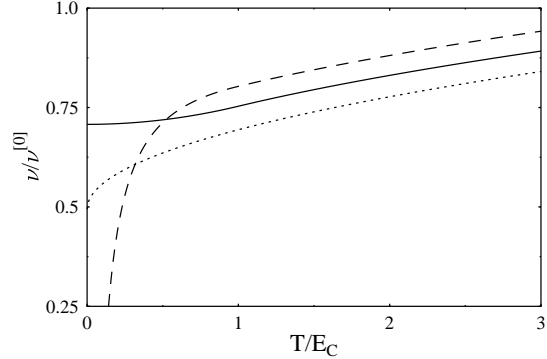}
\end{center}
\caption{\label{f2}   
Perturbative calculation of the DOS. 
Full line: ${\bf Q}\neq 0$ modes only; dashed 
line: ${\bf Q}= 0$ mode is included. Dotted line: AA result without account 
for  finite size effects ($d=3$, cf. Eq.~(6)). 
The curves are shifted vertically for clarity.}
\end{figure}

Let us consider a three dimensional cube with the linear size $L$ 
with no current flowing through the boundaries. 
The slow momentum ${\bf Q}$ assumes quantized values 
$${\bf Q}=\frac{2\pi}{L}{\bf n},$$
where ${\bf n}$ is a vector with integer components, including 
${\bf n}=(0,0,0)$. 
If either the energy, $\epsilon$, or the temperature, $T$, is much 
larger than the Thouless energy, $E_c\equiv D/L^2$, one may disregard the 
discreteness of  ${\bf Q}$ and perform slow momentum integration instead of 
summation. We shall refer to such a case as a three--dimensional, $d=3$. 
Substituting Eqs.~(\ref{scr}) and (\ref{pol}) in Eq.~(\ref{AA}) and 
performing integrations, 
one readily obtains \cite{Altshuler85} (e.g. for $T\gg E_c;\epsilon$)
\begin{equation}
\frac{\delta\nu_3}{\nu^{[0]}}=a_3\frac{1}{g}\sqrt{\frac{T}{E_c}},
                                                          \label{AA3}
\end{equation}
where $a_3\approx 3.8\cdot 10^{-2}$  and 
$g\equiv \nu^{[0]} D L^{d-2}\gg 1$ is the 
dimensionless conductance of the sample (the full $T$ and $\epsilon$ 
dependence may be found in Ref. \cite{Altshuler85}). 
At small temperature this correction 
exhibits singular (non--analytic) behavior, which is  the ZBA. 
One cannot employ, however, Eq.~(\ref{AA3}) for $\epsilon;T<E_c$. 
In this case the discreteness of the momentum spectrum begins to play a 
crucial role. Performing momentum summation in 
Eq.~(\ref{AA}), excluding the ${\bf Q}=0$ contribution, one obtains for 
$\gamma_{in}\ll \epsilon; T\ll E_c$ 
\begin{equation}
\frac{\delta\nu_3}{\nu^{[0]}}=
a_0\frac{1}{g}\left(\frac{T}{E_c}\right)^2 
f_0\left( \frac{\epsilon}{T} \right),  
                                                          \label{AA0}
\end{equation}
where $a_0=\sum_{{\bf n}\neq 0}(2\pi |{\bf n}|)^{-6}\approx 1.1\cdot 10^{-3}$ 
and 
$$f_0(x)=\int_0^\infty \!\! dy y\left( 
2-\mbox{tanh}\frac{y-x}{2}-\mbox{tanh}\frac{y+x}{2} \right) $$
with the asymptotic values $f_0(x\ll 1)=\pi^2/3\, $; $f_0(x\gg 1)=x^2$. 
At $T\approx E_c$ Eqs.~(\ref{AA3}) and (\ref{AA0})  match parametrically. 
According to 
Eq.~(\ref{AA0}) the small  temperature singularity in the DOS is rounded off 
due to finite size effects (the finite value of $E_c$), (Fig. \ref{f2}).

Eq.~(\ref{AA0}) does not account, however, for the ${\bf Q}=0$ contribution 
to the momentum sum in Eq.~(\ref{AA}). Below we  evaluate 
this  contribution and argue that it reflects the physics 
of the Coulomb blockade. The first problem is that the bare Coulomb 
interaction, $V^{[0]}({\bf Q})=L^{-3} 4\pi e^2/Q^2$, is not well--defined for 
${\bf Q}=0$. We argue that for finite size samples this expression should be 
regularized at $Q\approx 1/L$ (cf.  analogous regularization of the bare 
interaction in the $d=1$ case \cite{Altshuler85}), leading to 
$V^{[0]}({\bf Q}=0)\approx  e^2/L$. More precisely we shall use 
\begin{equation}
V^{[0]} = \frac{e^2}{C},
                                                \label{v00}
\end{equation}                                                
where $C\approx L$ is the {\em self} capacitance of the sample, 
which includes some 
non--universal geometrical factors (hereafter the interaction potential, $V$, 
without momentum index refers to ${\bf Q}=0$). 
In other words, one may regard the finite 
value of $V^{[0]}$  as the  result of fast screening 
processes, occurring on the boundary of the sample, which are not included 
explicitly in Eq.~(\ref{scr}). According to Eqs.~(\ref{scr}) and (\ref{pol}), 
the ${\bf Q}=0$ interaction can not be screened, indeed 
$\Pi({\bf Q}=0,\omega_m)=0$. This amounts to  saying that a finite size 
{\em isolated} system cannot screen its total charge. This is not the case 
for a system which is coupled capacitevely  to the environment. In the latter 
case the polarization operator should contain an additional term arising 
from the 
polarization of the environment. This would lead to an effective screening of the 
${\bf Q}=0$ interaction. We postpone, however, discussion of the screened 
zero--mode interaction till Section \ref{s4} and proceed with the 
interaction given by Eq.~(\ref{v00}). 

Substituting now Eq.~(\ref{v00}) into Eq.~(\ref{AA}) and performing energy 
summation and analytical continuation, one obtains for the zero--mode 
contribution to the DOS  
\begin{equation}
\frac{\delta\nu_0}{\nu^{[0]}}=-\frac{V^{[0]} }{2\pi^2  T}
\Re\,\Psi^{(\prime)}\left(\frac{1}{2}+
\frac{\gamma_{in}-i\epsilon}{2\pi T}\right), 
                                                          \label{AA1}
\end{equation}
where $\Psi(x)$ is  the digamma function. For 
$\gamma_{in}\ll T$ this reduces to 
\begin{equation}
\frac{\delta\nu_0}{\nu^{[0]}}=-\frac{V^{[0]}}{4 T}
\mbox{cosh}^{-2}\frac{\epsilon}{2 T}.  
                                                          \label{AA2}
\end{equation}
Note that the dependencies on temperature and energy are very different, 
whereas for finite $d$ energy and temperature play essentially the same role. 
The  temperature (and diffusive constant) dependence is compatible with the 
AA result,  
$\delta\nu_d\sim T^{(d/2-1)}D^{-d/2}$, extended to  $d=0$. The zero--mode 
contribution 
leads to a dramatic singularity $\sim T^{-1}$ in the single particle DOS, 
Fig.~\ref{f2}. 
For sufficiently low  temperatures, $T\ll e^2/2C$, 
this last expression cannot be correct. Indeed in this 
case Eq.~(\ref{AA2}) predicts $\delta\nu\gg\nu^{[0]}$, a result which is 
certainly nonperturbative. Moreover, the  resulting DOS 
becomes negative. This clearly indicates that the first order perturbation 
theory is not sufficient to treat the ZBA in 
finite size systems at low temperatures. 
In the next section we shall show how the zero--mode interaction can be 
treated nonperturbativly. 
Unlike Eq.~(\ref{AA2}), the exact result is  well 
behaved at any temperature and, in fact, is well known from the 
``orthodox'' theory of the Coulomb blockade \cite{Likharev89}.

\section{Zero mode interaction}
\label{s3}

\subsection{General formulation}
\label{s3a}

As we have seen in the previous section, the perturbative treatment of the 
zero--mode interactions in a dot at low temperature 
meets  serious difficulties. From another hand, the first order (in the 
screened interaction) result for 
${\bf Q} \neq 0$ and a not too dirty system, $g\gg 1$ (cf. 
Eq.~(\ref{AA3}), (\ref{AA0})), is well behaved. Thus we separate out the 
dangerous contribution of the zero--mode interaction and try to treat it 
non--perturbativly. To this end  consider  a Hamiltonian 
describing  electrons  moving in 
a disordered potential, which interact  through the zero mode 
{\em repulsive} interaction only 
\begin{equation}
H_0=\sum_{\alpha}\epsilon_{\alpha}a^{+}_{\alpha}a_{\alpha}+
\frac{V^{[0]}}{2}:\left[ \sum_{\alpha}a^{+}_{\alpha}a_{\alpha} -N_0 
\right]^2:\, , 
                                                            \label{H}
\end{equation}                                                            
where $a^{+}_{\alpha} (a_{\alpha})$ is a creation (annihilation) operator of 
an electron in an exact single particle state (which is defined including 
disorder potential and spin) with an eigenenergy $\epsilon_{\alpha}$; 
$V^{[0]}$ is the bare zero--mode interaction, Eq.~(\ref{v00}); 
$N_0$ - the  charge of the positive background, which we assume to be 
an integer \cite{foot3}.
Bellow we shall comment on the case where Hamiltonian includes also finite 
${\bf Q}$ interactions, which can be treated perturbatively. 

Before to proceed we would like to stress the following fact.  
One may argue that the interaction term in the 
Hamiltonian, Eq.~(\ref{H}), is trivial. Indeed it has the form 
$H_{int}= V^{[0]} :N^2:/2$, where 
$N\equiv \sum_{\alpha}a^{+}_{\alpha}a_{\alpha}$ is a total number of 
particles in the system. As $N$ commutes with the Hamiltonian, 
$[H_0,N]=0$, it does not 
have  any dynamics, $\partial_t N=0$. Thus the interaction term is just a 
constant added to the 
Hamiltonian, and seems not to have any nontrivial consequences. This is 
indeed the case when considering  thermodynamical properties or response 
functions of the isolated 
dot. It is, however, {\em not} the case with  the single 
particle Green 
function, which correlates the amplitude of the creation of one additional 
electron at an initial time, $t_i$, and its subsequent destruction at 
time $t_f$. Thus any measurement of the single 
particle Green function assumes implicitly the existence of processes which 
do not conserve the number of particles in a dot (e.g. due to tunneling). 
Such processes 
render the entire Hamiltonian noncommuting with the particle number. As a 
result the $N^2$ interaction term has a significant impact on the single 
particle Green function. After completing the calculations 
we shall comment on the relation of the above arguments to gauge invariance.

The  
imaginary time single particle Green function may be written as \cite{Negele} 
\begin{equation}
{\cal G}_{\alpha}(\tau_i,\tau_f,\mu)=\frac{1}{Z(\mu)}
\int{\cal D}[\psi^{*}_{\alpha}(\tau)\psi_{\alpha}(\tau)]
e^{-S[\psi^{*}_{\alpha},\psi_{\alpha}]  }
\psi^{*}_{\alpha}(\tau_i)\psi_{\alpha}(\tau_f), 
                                                         \label{G}
\end{equation} 
with the fermionic action  given by 
\begin{equation}
S[\psi^{*}_{\alpha},\psi_{\alpha}]= 
\int_0^{\beta}d\tau \left[ \sum_{\alpha}
\psi^{*}_{\alpha}(\tau) 
(\partial_{\tau}+\epsilon_\alpha-\mu) \psi_{\alpha}(\tau) 
+\frac{V^{[0]}}{2} 
\left[ \sum_{\alpha}\psi^{*}_{\alpha}(\tau)\psi_{\alpha}(\tau) -N_0\right]^2 
\right ]; 
                                                         \label{action}
\end{equation}
here $Z(\mu)$ is the partition function and $\mu$ is the chemical potential. 
Splitting the interaction term in the  action  by means of the 
Hubbard--Stratonovich transformation with 
the  auxiliary Bose field, $\phi(\tau)$, one obtains  
\begin{eqnarray}
{\cal G}_{\alpha}(\tau_i-\tau_f)=\frac{1}{Z(\mu)}
&&\int{\cal D}[\phi(\tau)]
e^{-\int_0^{\beta}d\tau  \left[ 
\phi(\tau)[2V^{[0]}]^{-1}\phi(\tau) -i N_0\phi(\tau) \right] }  \nonumber \\
&&\int{\cal D}[\psi^{*}_{\alpha}\psi_{\alpha}]
e^{-\int_0^{\beta}d\tau 
\sum_{\alpha}\psi^{*}_{\alpha} 
(\partial_{\tau}+\epsilon_{\alpha}-\mu+i\phi(\tau)) \psi_{\alpha} }
\psi^{*}_{\alpha}(\tau_i)\psi_{\alpha}(\tau_f)     \nonumber \\
=\frac{1}{Z(\mu)}&&\int{\cal D}[\phi(\tau)]
e^{-\int_0^{\beta}d\tau  \left[ 
\phi(\tau)[2V^{[0]}]^{-1}\phi(\tau) -i N_0\phi(\tau) \right] }
Z^{[\phi]}(\mu){\cal G}^{[\phi]}_{\alpha}(\tau_i,\tau_f,\mu)
                                                          \label{HS}
\end{eqnarray} 
with the same transformations in $Z(\mu)$. Here $Z^{[\phi]}(\mu)$ and 
${\cal G}^{[\phi]}_{\alpha}(\tau_i,\tau_f,\mu)$ are  respectively 
the partition  
and  Green functions of non--interacting electrons in the time 
dependent (but spatially uniform) potential, 
$i\phi(\tau)$. To calculate these quantities one should resolve  
the spectral problem for the first order differential operator 
$\left[ \partial_{\tau}-\mu+\epsilon_{\alpha}+i\phi(\tau) \right]$ 
with  antiperiodic 
boundary conditions. This  can be easily done (see e.g. chapter  7 in 
Ref. \cite{Negele}). For the spectral determinant (partition function) one 
finds 
\begin{equation}
Z^{[\phi]}(\mu)=Z^{[0]}(\mu-i\phi_0),
                                                          \label{ZG1}
\end{equation}
where $Z^{[0]}(\mu)\equiv\exp\{-\beta\Omega^{[0]}(\mu)\}$ is the partition 
function  of non--interacting electron gas. We have introduced Matsubara 
representation for the boson field, $\phi(\tau)$: 
\mbox{$\phi_m\equiv
\beta^{-1}\int_0^{\beta}d\tau\phi(\tau)\exp\{i\omega_m\tau\}$}, 
$\omega_m=2\pi m T$. 
The Green function is given by 
\begin{equation}
{\cal G}^{[\phi]}_{\alpha}(\tau_i,\tau_f,\mu)=
{\cal G}^{[0]}_{\alpha}(\tau_i-\tau_f,\mu-i\phi_0)
e^{i\int_{\tau_i}^{\tau_f}d\tau [\phi(\tau)-\phi_0]}, 
                                                          \label{ZG}
\end{equation} 
where ${\cal G}^{[0]}_{\alpha}(\epsilon_n,\mu)=
(i\epsilon_n-\epsilon_{\alpha}+\mu)^{-1}$ 
is the  Green function of  non-interacting fermions. 
It is convenient to rewrite 
the exponent in the last equation in the following form 
\begin{equation}
\exp\{i\int_{\tau_i}^{\tau_f}d\tau [\phi(\tau)-\phi_0]\} =
\exp\{\beta\sum_{m\neq 0}\frac{\phi_m J_{-m}^{\tau_i,\tau_f}}{\omega_m} \}, 
                                                          \label{exp}
\end{equation} 
where $J_m^{\tau_i,\tau_f}$ is the Matsubara transform of the following 
function 
\begin{equation}
J_{\tau}^{\tau_i,\tau_f}=\delta(\tau-\tau_i)-\delta(\tau-\tau_f).  
                                                          \label{J}
\end{equation} 

\begin{figure}[htbp]
\epsfysize=3cm
\begin{center}
\leavevmode
\epsfbox{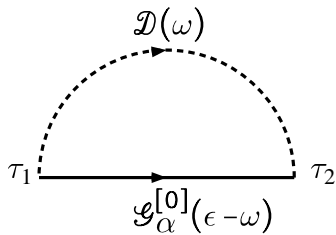}
\end{center}
\caption{\label{f3} 
Diagrammatic representation of the Coulomb boson.}
\end{figure}

Transforming next 
the functional integral over $\phi(\tau)$ to integrals 
over the Matsubara components, $\phi_m$, we obtain 
\begin{eqnarray}
{\cal G}_{\alpha}(\tau_i-\tau_f)=\frac{1}{Z(\mu)}
&&\int d\phi_0
e^{-\beta[\phi_0[2V^{[0]}]^{-1}\phi_0 -i\phi_0 N_0+\Omega^{[0]}(\mu-i\phi_0)]}
{\cal G}^{[0]}_{\alpha}(\tau_i-\tau_f,\mu-i\phi_0)    \nonumber  \\
&&\int\prod_{m\neq 0} d\phi_m
\exp \left\{ \beta  \sum_{m\neq 0} \left[ -\frac{\phi_m\phi_{-m}}{2V^{[0]}}
+\frac{\phi_m J_{-m}^{\tau_i,\tau_f}}{\omega_m} \right] \right\}. 
                                                          \label{G1}
\end{eqnarray} 
Again with the analogous modifications in $Z(\mu)$. 
The integral over the static component, $\phi_0$, describes the smooth 
transition between the grand canonical ensemble with the chemical potential 
$\mu$ (at $V^{[0]}=0$) to the canonical ensemble with $N_0$   
electrons (at $V^{[0]}=\infty$). For   large enough systems 
($\Delta^{-1}\equiv-\partial^2\Omega^{[0]}/\partial\mu^2\gg \beta$) one can neglect 
differences between the two statistical ensembles. This means that the integral 
over $\phi_0$ can be calculated in a saddle point approximation, leading to 
${\cal G}^{[0]}_{\alpha}(\tau_i-\tau_f,\overline\mu)$, where 
the stationary point, $\overline\mu$, is the real  solution of the equation 
$(\mu-\overline\mu)/V^{[0]}+N_0+\partial \Omega^{[0]}(\overline\mu)
/\partial\mu=0$. 
The remaining  integrals (over $\phi_m$ for $m\neq 0$) are purely Gaussian. 
As a result one obtains 
\begin{equation}
{\cal G}_{\alpha}(\tau_i-\tau_f,\mu)=
{\cal G}^{[0]}_{\alpha}(\tau_i-\tau_f,\overline\mu)
e^{-S(\tau_i-\tau_f)}, 
                                                          \label{GT}
\end{equation} 
where 
\begin{equation}
S(\tau)=T\sum_{m\neq 0}\frac{V^{[0]}}{\omega_m^2}
(1-e^{i\omega_m\tau}). 
                                                          \label{S}
\end{equation} 
Eqs.~(\ref{GT}) and (\ref{S}) solve the problem of finding the exact single 
particle Green function in the presence of the zero--mode interaction. 
This result is depicted diagrammatically in  Fig. \ref{f3}. The dressed 
Green function of the interacting problem is given by the bare one (with a
renormalized chemical potential) 
decorated with the propagator of the auxiliary boson field 
${\cal D}(\tau)\equiv \exp\{-S(\tau)\}$. 
We refer to this auxiliary field as a  -- {\em Coulomb boson}. 
As ${\cal D}(0)=1$, the zero mode interaction 
does not influence equal time Green functions (apart from the renormalization of 
the chemical potential). As a result thermodynamical quantities are not 
affected by the presence of the Coulomb boson. This is not unexpected since 
the interaction term commutes with a total Hamiltonian, hence it does not 
affect  
quantities defined for the closed system. By contrast, to measure a 
two point Green function one should perform a tunneling experiment, where 
the system can not be considered as completely isolated. In this case the 
presence of the interaction term in Eq.~(\ref{H}) 
(and hence of the Coulomb boson) is of crucial importance.

This observation can be related to  gauge invariance \cite{Fin}.
As we have seen the problem with zero mode interaction is essentially 
reducible to that of an electrons gas in  a spatially uniform a.c. potential. 
Such a 
potential can be always removed from the problem by a time--dependent gauge 
transformation. Thus gauge invariant physical quantities (e.g. 
thermodynamical quantities) are not affected by 
the presence of spatially uniform potential, cf. Eq.~(\ref{ZG1}). 
The single particle Green function, being 
{\em non} gauge invariant object is affected. 
By introducing an electron tunneling from an external source we fix the (time 
dependent) phase of the electron wave function in the system. This point has 
been discussed by Finkelstein \cite{Fin}, who argued that unlike the 
conductivity or  thermodynamical quantities, the non gauge invariant single 
particle DOS may be affected by very small ${\bf Q}$ terms, leading in $d\leq 
2$ to more pronounced singularities.

\subsection{The Coulomb boson and DOS} 
\label{s3b}

\begin{figure}[htbp]
\epsfysize=8cm
\begin{center}
\leavevmode
\epsfbox{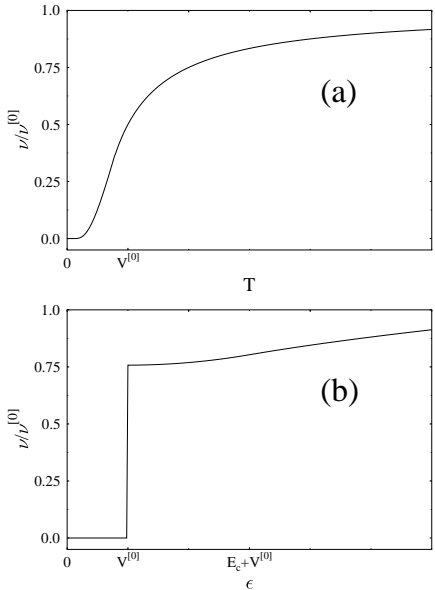}
\end{center}
\caption{\label{f4}  
Non perturbative 
DOS. (a) as function of temperature at the Fermi energy 
($\epsilon=0$); only the ${\bf Q}= 0$ mode is included. 
(b) DOS as function of energy at $T=0$; finite ${\bf Q}$ contributions 
are included.}
\end{figure}

For the instantaneous (frequency--independent) interaction, Eq.~(\ref{v00}), 
the sum in Eq.~(\ref{S}) may be easily performed, yielding 
\begin{equation}
S(\tau)=\frac{V^{[0]}}{2}
\left( |\tau|-\frac{\tau^2}{\beta} \right). 
                                                          \label{S0}
\end{equation} 
Using the Lehmann representation for  temperature Green functions 
\cite{Abrikosov64} one may write for the propagator of the Coulomb boson 
\begin{equation}
{\cal D}(\omega_m)=\int_{-\infty}^{\infty} \frac{d\omega^\prime}{2\pi}
\frac{B(\omega^\prime)}{i\omega_m-\omega^\prime}, 
                                                          \label{DL}
\end{equation}
where the spectral function  $B(\omega)$ is defined as  
$B(\omega)\equiv -2\Im{\cal D}^R(\omega)$. Fourier transforming  
of ${\cal D}(\tau)=\exp\{-S(\tau)\}$ and analytically continuing we obtain   
(see Appendix \ref{app}) 
\begin{equation}
B(\omega)=\sqrt{\frac{2\pi}{V^{[0]}T}} \left(
e^{-\frac{(\omega+V^{[0]}/2)^2}{2V^{[0]}T} }- 
e^{-\frac{(\omega-V^{[0]}/2)^2}{2V^{[0]}T} } \right). 
                                                          \label{B1}
\end{equation}
According to Eq. (\ref{GT}) (cf. also Fig. \ref{f3}) the 
electron Green function is given by 
\mbox{${\cal G}_{\alpha}(\epsilon_n)=T\sum_{\omega_m}
{\cal G}^{[0]}_{\alpha}(\epsilon_n-\omega_m)
{\cal D}(\omega_m)$}. Performing the summation by a  standard contour 
integration and  then 
analytical continuation ($i\epsilon_n\rightarrow\epsilon+i\delta$) 
one obtains  for the one particle 
(tunneling) density of states, $\nu(\epsilon)\equiv-\pi^{-1}\sum_{\alpha}\Im
{\cal G}^{R}_{\alpha}(\epsilon)$, 
\begin{equation}
\nu(\epsilon)=-\frac{1}{2}\int_{-\infty}^{\infty} 
\frac{d\omega}{2\pi}
\left( 
\mbox{tanh}\frac{\epsilon-\omega}{2T}+  
\mbox{coth}\frac{\omega}{2T} 
\right) 
B(\omega) \nu^{[0]}(\epsilon-\omega),
                                                          \label{nu}
\end{equation}
where $\nu^{[0]}(\epsilon)$ is the density of states in the absence of the 
zero--mode interaction and $B(\omega)$ is given by Eq.~(\ref{B1}). 
For  non--interacting electrons (with or without disorder) 
$\nu^{[0]}(\epsilon)$ may be well--approximated by a constant value, 
$\nu^{[0]}$, 
in this case the integral in the last expression may be readily evaluated. 
In the limit of  weak interaction ($V^{[0]}\ll T$) one obtains 
\begin{equation}
\frac{\nu_0}{\nu^{[0]}} =1-\frac{V^{[0]}}{4T}
\mbox{cosh}^{-2}\frac{\epsilon}{2T}. 
                                                          \label{nul}
\end{equation}
This is  the zero bias anomaly of AA extrapolated to  $d=0$ 
system, which has been obtained from a diagrammatic expansion, Eq. (\ref{AA2}). 
Notice that we did not actually use the fact that the 
system is disordered. Note also, that the exact lowest order result, 
Eq. (\ref{nul}), coincides with the one obtained from the exchange diagram 
only, Eq. (\ref{AA2}). The first order Hartree term with the ${\bf Q}=0$ 
interaction leads to  redefinition of the chemical potential which has been 
absorbed in the factor, $\overline \mu$. 
For  strong interaction ($V^{[0]}\gg T$) one has (Fig. \ref{f4}a)
\begin{equation}
\frac{\nu_0}{\nu^{[0]}} =\left\{ \begin{array}{ll} 
{\displaystyle \sqrt{\frac{2\pi T}{V^{[0]}}} \,
e^{-\frac{V^{[0]}}{8T}} \mbox{cosh}\frac{\epsilon}{2T}  }; \,\,\,\, 
&\epsilon\ll \sqrt{V^{[0]}T} \ll V^{[0]}, \\
{\displaystyle 1-e^{-(\epsilon-V^{[0]})/T }  };   
& \epsilon\gg V^{[0]}\gg T.
\end{array} \right. 
                                                          \label{nug}
\end{equation}
The exponential suppression of the tunneling density of states near the 
Fermi energy is a direct manifestation of the Coulomb blockade. The ZBA 
in $d=0$ systems, being treated non--perturbativly, leads to  results which 
are well--known from the ``orthodox'' theory of the Coulomb blockade, cf. Ref. 
\cite{Likharev89}. 

One may ask how the finite ${\bf Q}$ interactions modify the standard 
Coulomb blockade predictions. 
In the case where the finite ${\bf Q}$ interactions can be treated 
perturbativly the answer can be read off Eq. (\ref{nu}): in  Eq. (\ref{nu}) 
one should  substitute the perturbative AA expression for the 
DOS ($\nu^{AA}$) for  $\nu^{[0]}(\epsilon)$. 
Indeed, one may repeat 
the  derivation given above in the presence of the finite ${\bf Q}$ 
interactions; 
the result coincides  with  Eq. (\ref{GT}), where ${\cal G}^{[0]}$  includes 
the effects of all but zero--mode interactions. 
In the limit  of zero temperature  one  obtains 
\begin{equation}
\nu(\epsilon) =\left\{ \begin{array}{ll} 
{\displaystyle 0}; &|\epsilon|< V^{[0]}/2,  \\
{\displaystyle \nu^{AA}(|\epsilon|-V^{[0]}/2)} ;\,\,\,\, 
   & |\epsilon| > V^{[0]}/2. 
\end{array} \right. 
                                                          \label{nui}
\end{equation}
This result  (Fig. \ref{f4}b) implies that for voltages larger than the Coulomb 
blockade gap the tunneling current is still somewhat suppressed due to the 
presence of the 
finite dimensional ZBA. Although the finite ${\bf Q}$ interactions 
reduce the jump at $|\epsilon| = V^{[0]}/2$ ($T=0$), it  do not remove the 
discontinuity. There is, however, important physics which leads to 
the rounding  off  of the threshold at $|\epsilon|=e^2/2C$ even at zero 
temperature. 
This is the  screening of the zero--mode interaction due to fluctuations in 
the electromagnetic environment. This issue is to be discussed next.

\section{Screened  zero--mode interaction} 
\label{s4}

In the preceding section we  studied the tunneling DOS of a dot which is 
perfectly  isolated (both electrically and electromagnetically) from the 
outside world. In this case the total charge of the dot can not be screened. 
This is why we have employed a bare (instantaneous) zero--mode interaction, 
Eq. (\ref{v00}). This scheme, however, is not very satisfactory. First, 
measurements of some 
quantities of major interest (such as the single particle DOS studied here, 
or the inelastic broadening to be discussed elsewhere), require by 
definition direct coupling to an external medium. But more importantly, even in 
the absence of external leads the dot interacts (through capacitive coupling) 
with external gates, conducting layers, external charges, all of which will 
be referred to as the environment. The creation of an  
additional charge in the dot leads to redistribution of charges in the 
environment, which takes  finite time. The redistributed 
environmental charge  partially screens the initially created charge on the 
dot, reducing the zero--mode interaction energy. The fact that this 
redistribution is not instantaneous implies that the effective screened 
interaction is retarded. Equivalently one may notice  that 
the dot's capacitance  is proportional to the dielectric constant of the 
surrounding medium. Unless this medium is a perfect vacuum, 
the dielectric constant is 
a function of frequency. As a result the zero--mode 
interaction, $e^2/C(\omega)$, is not instantaneous, but rather 
characterized by a finite retardation time scale. 

\begin{figure}[htbp]
\epsfysize=4cm
\begin{center}
\leavevmode
\epsfbox{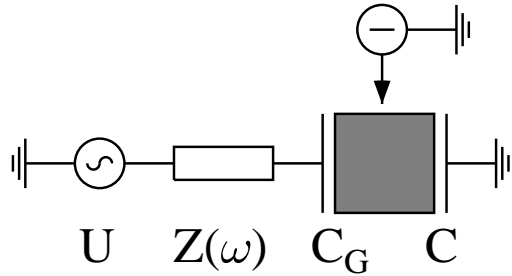}
\end{center}
\caption{\label{f5}   
Equivalent circuit of a dot coupled to the enviroment. 
The grey square represents  a dot, $C$ being its self  capacitance; 
$C_G$-- the mutual capacitnce to the environment; 
$Z(\omega)$-- the linear impedance of the environment. Here $U$ represents 
the noise voltage. The black arrow is a tunneling tip through which particles 
may be injected to the dot, measuring its  DOS. This tip may be coupled to 
the dot through a large impedance barrier. }
\end{figure} 

Phenomenologically it is convenient to consider an equivalent circuit 
(Fig. \ref{f5}), assigning the self capacitance, $C$, (bare interaction) to the 
dot, which in turn is capacitively coupled (through $C_G$) to the  gate 
electrode. The gate  is electrically coupled 
to the ground through the linear impedance $Z(\omega)$. The voltage source, 
$U$, represents the equilibrium noise voltage of the entire circuit.  
We shall model the dot--environment interaction by assuming that the dot is 
subjected  to a  time 
dependent (but spatially uniform) noise potential due to  the environment 
\begin{equation}
H_{noise}=\eta(t)\sum_{\alpha}a^{+}_{\alpha}a_{\alpha}.
                                                          \label{noise}
\end{equation}
At the end of these calculations all physical quantities should be averaged over 
realizations of the noise. We shall assume further that the noise is Gaussian 
with zero mean value. The averaging procedure thus may be 
written as 
\begin{equation}
\langle \ldots \rangle_{noise} =\frac{1}{{\cal N}}
\int{\cal D}[\eta(\tau)]
e^{-\frac{1}{2}\int\int_0^\beta d\tau d\tau^{\prime} 
\eta(\tau)K^{-1}(\tau-\tau^{\prime})\eta(\tau^{\prime})  }
\ldots\,\, , 
                                                         \label{aver}
\end{equation} 
where ${\cal N}$ is a normalization factor and 
$K(\tau-\tau^{\prime})=\langle \eta(\tau) \eta(\tau^{\prime}) \rangle$ is 
the noise correlator, which is to be determined employing the 
fluctuation--dissipation theorem (FDT). 
An important observation is  that the partition 
function of the dot, $Z(\mu)$, is not affected by the noise term. Indeed, we 
have already seen, Eq.~(\ref{ZG1}), that the partition function of the 
electron gas in a spatially uniform a.c. field depends on its mean value 
only (which is zero in our case). This 
can be understood as a consequence of the gauge invariance of the partition 
function -- spatially uniform field may be always removed by a gauge 
transformation. As a result,  the noise term  in Eq.~(\ref{G}) enters  in the  
numerator only. Averaging Eq.~(\ref{G}) over 
the Gaussian noise, Eq.~(\ref{aver}), leads to an effective fermionic action 
with  interaction which is  non--local in time, 
\begin{equation}
S_{int}[\psi^{*}_{\alpha},\psi_{\alpha}]= \frac{1}{2}
\int \!\!\! \int_0^{\beta}d\tau d\tau^{\prime}  
\sum_{\alpha}\psi^{*}_{\alpha}(\tau)\psi_{\alpha}(\tau) 
\left( V^{[0]}\delta(\tau -\tau^{\prime}) - K(\tau-\tau^{\prime}) \right)  
\sum_{\alpha}\psi^{*}_{\alpha}(\tau^{\prime})\psi_{\alpha}(\tau^{\prime}) .
                                                         \label{nlact}
\end{equation}
As a result, one obtains an effective renormalization (screening) 
of the zero--mode interaction potential 
\begin{equation} 
V^{[0]} \rightarrow V(\omega_m)=V^{[0]}- K(\omega_m).
                                                          \label{Veff}
\end{equation}                        
Further calculations follow  the same steps outlined in Sec. \ref{s2}. 
The final result is given 
again by Eq.~(\ref{GT}) where now (cf.   Eq.~(\ref{S})) 
\begin{equation}
S(\tau)=T\sum_{m\neq 0}\frac{V(\omega_m)}{\omega_m^2}
(1-e^{i\omega_m\tau}). 
                                                          \label{Sscr}
\end{equation} 

We shall  next employ the FDT to calculate the  noise correlator, 
$K(i\omega)$. According to the FDT the equilibrium noise spectrum ($T=0$) 
of the total noise voltage generated by the circuit is 
$\langle UU\rangle=e^2 i\omega Z_{tot}$, where the total impedance of the 
equivalent circuit is 
$Z_{tot}=(i\omega C)^{-1}+(i\omega C_G)^{-1}+Z(\omega)$. 
The corresponding voltage drop on the dot is $\eta=U(i\omega C)^{-1}/Z_{tot}$; 
thus the noise correlator, $K=\langle \eta\eta \rangle$, 
is given by
\begin{equation} 
K(i\omega)=\frac{e^2}{C}\frac{1/(i\omega C)}{Z_{tot}}. 
                                                          \label{K}
\end{equation}
Substituting this expression in  Eq.~(\ref{Veff}) and rewriting it in a 
finite temperature form one obtains
\begin{equation}
V(\omega_m)=\frac{e^2}{C}\, \, 
\frac{C/C_G+|\omega_m| ZC}{1+C/C_G+|\omega_m| ZC}. 
                                                          \label{v0scr}
\end{equation}
The high frequency (unscreened) value coincides with that of Eq.~(\ref{v00}), 
whereas at low frequency the interaction is partially screened and is 
given by the total capacitance, $V(0)=e^2/(C+C_G)$. 
The characteristic crossover frequency 
is given by the ``$RC$'' time of the circuit, $\omega\approx (ZC)^{-1}$. 
As a result the long time  behavior of $S(\tau)$,  
determined by the small frequency asymptotic  of the screened interaction, 
is given by 
$S(\tau)\approx |\tau| e^2/2(C+C_G) $ for $ZC\ll\tau\ll\beta$. 

A case of particular interest is that of a ``maximally'' screened interaction, 
when \mbox{$V(i\omega\rightarrow0)=0$}. 
In this case the linear term in $S(\tau)$ is absent and 
the action grows at most logarithmically at large $\tau$. 
This would immediately imply that instead 
of exponential suppression of the DOS at the Fermi energy, one obtains only 
a power--law ZBA. According to Eq.~(\ref{v0scr}) such a case may be realized
when $C_G\rightarrow \infty$ (more precisely 
$ZC_G\gg\beta$). This is the case of a dot strongly connected to the 
environment. For simplicity we restrict ourselves to  the scenario of a pure 
ohmic 
environment, $Z=R$, in which case the interaction potential is given by 
\cite{foot4} 
\begin{equation}
V(\omega_m)=V^{[0]}\, \frac{|\omega_m|}{\Omega+|\omega_m|}, 
                                                          \label{Vstr}
\end{equation} 
where $\Omega\equiv (RC)^{-1}$ and $V^{[0]}$ is given by Eq.~(\ref{v00}). 
As  $V(\omega=0)=0$ the addition or subtraction of an electron from the dot 
costs no Coulomb energy over long time scales. 
On  short time scales, however, before the environment adjusts to screen out 
the added electron, one has to pay some energy, Eq. (\ref{Vstr}). Thus the 
addition of an electron can be considered as  ``tunneling'' under an energy 
barrier in the time direction. The very  same physics has been discussed  in 
the context of $d=2$ systems in Refs. \cite{Spivak94,Levitov95}. 
The related energy cost on short time scales suppresses free  particle exchange 
between the dot and the particle reservoir (although 
Coulomb blockade in its strict sense is absent). This leads to the 
suppression of the tunneling DOS hence to ZBA.
Substituting Eq. (\ref{Vstr}) into Eq. (\ref{Sscr}), one obtains 
for  $\Omega^{-1}\ll \tau\ll \beta$ \cite{foot5} 
\begin{equation} 
S(\tau)\approx 2r\log(1+\tilde\Omega|\tau|)+
O\left( (\Omega\tau)^{-1} \right)\, ,
                                                      \label{sas}
\end{equation} 
where $\tilde\Omega\equiv e^{\gamma}\Omega$ with $\gamma=0.577\ldots$ 
being the Euler constant and 
$r\equiv V^{[0]}/(2\pi\Omega)= R e^2/(2\pi \hbar)$ 
is dimensionless resistance of the environment. 
Performing the Fourier transform and analytical 
continuation (see Appendix \ref{app}), 
one obtains for the zero--temperature spectral density of the Coulomb boson 
\begin{equation} 
B(\omega)=-\frac{\mbox{sign}\,\omega}{\tilde\Omega}\,
\frac{2\pi}{\Gamma(2r)} 
\left| \frac{\omega}{\tilde\Omega}  \right|^{2r-1}
e^{-|\omega/\tilde\Omega|}. 
                                                      \label{B0scr}
\end{equation} 
Finally the zero temperature DOS may be found from Eq.~(\ref{nu}) with 
$B(\omega)$ given by Eq.~(\ref{B0scr}). Assuming a constant bare DOS 
(i.e. $\nu^{[0]}=\mbox{const}$) one obtains 
\begin{equation}
\frac{\nu_0(\epsilon)}{\nu^{[0]}} =
1-\Gamma\left( 2r,\left| \frac{\epsilon}{\tilde\Omega}  \right|\right)\approx
\left\{ \begin{array}{ll} 
{\displaystyle \frac{1}{\Gamma(2r+1)} 
\left| \frac{\epsilon}{\tilde\Omega}  \right|^{2r}  }; 
&|\epsilon|\ll \Omega, \\
{\displaystyle 
1-\left| \frac{\epsilon}{\tilde\Omega}  \right|^{2r-1}
e^{ -|\epsilon/\tilde\Omega|}  };\,\,\,\,    
& |\epsilon|\gg \mbox{max}\{\Omega, V^{[0]}\}, 
\end{array} \right.
                                                          \label{nu0scr}
\end{equation}
where $\Gamma(2r,x)$ is the incomplete gamma function. 
For the maximally screened zero--mode interaction scenario the ZBA in $d=0$ 
has a power law behavior, rather than the gap given by 
Eq.~(\ref{nui}). For 
high--impedance (slow) environment, $r\gg 1$, there is a crossover to the 
``orthodox'' Coulomb blockade in the interval $\Omega<\epsilon<V^{[0]}$.
However for the low--impedance, $r<1$, environment the power--law ZBA, 
Eq.~(\ref{nu0scr}), directly  crosses over to the finite dimensional AA 
result at $\epsilon\approx \Omega$. 

The straightforward finite temperature generalization of Eq.~(\ref{B0scr}) 
is (see Appendix \ref{app})
\begin{equation} 
B(\omega)=-\frac{\mbox{sinh}\, \omega/2T}{\tilde\Omega}\,
\frac{2}{\Gamma(2r)} 
\left( \frac{2\pi T}{\tilde\Omega}  \right)^{2r-1}
\left| \Gamma\left( r+\frac{i\omega}{2\pi T} \right) \right|^2 . 
                                                      \label{Bscr}
\end{equation} 
Substituting this result into Eq.~(\ref{nu}) one obtains e.g. for the DOS at 
the Fermi energy, $\epsilon=0;\,\,\, T\ll\Omega$ 
\begin{equation}
\frac{\nu_0}{\nu^{[0]}} =a(r) 
\left(  \frac{2\pi T}{\tilde\Omega} \right)^{2r},  
                                                          \label{nuscr}
\end{equation}
where 
\begin{equation}
a(r)\equiv \frac{1}{\pi\Gamma(2r)}\int_0^\infty dx
\frac{\left| \Gamma( r+ix ) \right|^2}{\mbox{cosh}\, \pi x}=
\left\{ \begin{array}{ll}
1-\Psi\left( \frac{1}{2} \right)2r;\,\,\, & r\ll 1, \\
2^{-2r}/\sqrt{\pi r}; & r\gg 1;
\end{array} \right. 
                                                  \label{sigma}
\end{equation}
$\sigma(1/2)=1/\pi;\,\,\, \sigma(1)=1/8$. At 
$T > \mbox{max} \{\Omega, V^{[0]} \}$ the DOS obeys Eq.~(\ref{nul}). 
Screening of the zero mode 
interaction converts the  exponential suppression of the DOS  
(cf. Eq.~(\ref{nug})) into the power--law, Eq.~(\ref{nuscr}). 

Our results in the limit of maximally screened interaction $C_G\rightarrow 
\infty$ are compatible with 
the results of 
Refs.  \cite{Devoret90,Glazman90}, where  tunneling through a {\em single} 
tunnel junction coupled to a  linear impedance was considered. 
This coincidence is not surprising at all; indeed a dot coupled to a  
circuit through 
large capacitance, $C_G$, can be viewed as an island from which charge can 
continuously leak out; this is practically the setup of 
Refs.  \cite{Devoret90,Glazman90}. 
Our present analysis stresses  the physics of a weakly coupled dot 
(with a finite $C_G$, Eq. (\ref{v0scr})) and its 
relation to the ZBA.  At $T=0$, employing the above given expressions one 
obtains  
\begin{equation}
\frac{\nu_0(\epsilon)}{\nu^{[0]}}= 
\left\{ \begin{array}{ll} 
0;  & |\epsilon| \leq e^2/2(C+C_G), \\
1-\Gamma \left( 2\tilde r,
\frac{|\epsilon| - e^2/(C+C_G)}{\tilde\Omega(C+C_G)/C_G} \right); 
\,\,\,\, & e^2/2(C+C_G)< |\epsilon|, 
\end{array} \right.
                                                          \label{nu0ss}
\end{equation}
where $\tilde r\equiv r[C_G/(C+C_G)]^2$. 
At finite temperature and $\epsilon=0$ one has 
\begin{equation}
\frac{\nu_0}{\nu^{[0]}} \approx 
\left\{ \begin{array}{ll} 
\exp\{-e^2/8T(C+C_G)\}; \,\,\,\,  &  T \leq e^2/2(C+C_G), \\
(T/\tilde\Omega)^{2\tilde r}  ; &e^2/2(C+C_G)< T < \Omega, \\
1-e^2/4TC;   & \mbox{max}\{\Omega, e^2/2C\}< T 
\end{array} \right.
                                                          \label{nu0sss}
\end{equation}
(the above expressions account for the ${\bf Q}=0$ contribution only). 
We stress that, our  results are a particular, 
d=0, case of a general non--perturbative expression for the ZBA obtained by 
Nazarov \cite{Nazarov89} and Levitov and Shytov \cite{Levitov95}. Our point 
here is that for $d=0$ case all  calculations can be carried out exactly, 
avoiding some of the uncontrolled albeit plausible assumptions employed 
in Refs. \cite{Nazarov89,Levitov95}.

\section{Acknowledgments}

We are grateful to A. Altland,  M.~Devoret, U. Sivan and 
B. Shklovskii 
for  useful suggestions. In particular we are indebted to A. M. Finkelstein 
for his comments on the relation between the structure of the density of 
states and gauge invariance. This research was supported by the German--Israel 
Foundation (GIF) and the U.S.--Israel Binational Science Foundation (BSF) and 
the Israel Academy of Sciences.

\appendix
\section{}
\label{app}

\begin{figure}[htbp]
\epsfysize=5cm
\begin{center}
\leavevmode
\epsfbox{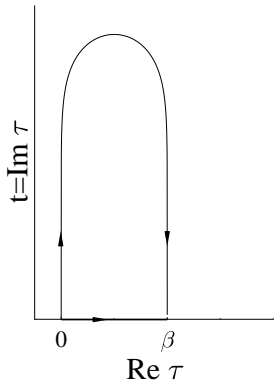}
\end{center}
\caption{\label{f6}    
Countour of integration in Eq.~(A1).}
\end{figure}

This Appendix is devoted to the calculation of the spectral density of the 
Coulomb boson, ${\cal D}(\tau)=\exp\{-S(\tau)\}$. In the case of unscreened 
zero mode interaction $S(\tau)$ is given by Eq.~(\ref{S0}). The Fourier 
transform has a form 
\begin{equation}
{\cal D}(\omega_m)=\int_0^\beta d\tau e^{i\omega_m\tau} 
e^{-\frac{V^{[0]}}{2}(\tau-\beta^{-1}\tau^2)} .
                                     \label{A1}
\end{equation}                                     
Deforming the contour of integration as it shown on Fig. \ref{f6} and making 
the obvious redefinition of variables one obtains 
\begin{equation}
{\cal D}(\omega_m)=i\int_0^\infty dt 
e^{-\omega_m t-\frac{V^{[0]}t^2}{2\beta}}
\left(  e^{-iV^{[0]}t/2} -  e^{iV^{[0]}t/2} \right).
                                     \label{A2}
\end{equation}    
In this expression one can perform analytical continuation, resulting in  
${\cal D}^R(\omega)={\cal D}(\omega_m\rightarrow -i\omega+\delta)$. The 
imaginary part of ${\cal D}^R(\omega)$ may be easily evaluated extending the 
region of integration up to $-\infty$. As a result, 
$B(\omega)=-2\Im{\cal D}^R(\omega)$ is given by Eq.~(\ref{B1}). 

Next we discuss the maximally screened zero mode interaction. For $0<\tau\ll 
\beta/2$ the action is given by Eq.~(\ref{sas}). Periodicity and symmetry 
requires that for $\beta/2\ll\tau< \beta$ action  has the  form 
$S(\tau)\approx 2r\log(1+\tilde\Omega(\beta-\tau))$.
Performing the Fourier transform by deforming the contour of integration and 
analytical continuation exactly as for the non--screened interaction, one obtains
$B(\omega)=-F(\omega)+F(-\omega)$, where 
\begin{equation} 
F(\omega)=2\Re\int_{0}^{\infty}dt
\frac{e^{i\omega t}}{(1+i\tilde\Omega t)^{2r}}=
\frac{\theta(\omega)}{\tilde\Omega}\,
\frac{2\pi}{\Gamma(2r)} 
\left( \frac{\omega}{\tilde\Omega}  \right)^{2r-1}
e^{-\omega/\tilde\Omega}. 
                                                      \label{A4}
\end{equation}
This leads to zero temperature expression, Eq.~(\ref{B0scr}). To generalize 
the above expressions to  finite temperatures we use the conformal 
transformation $\tau\rightarrow \sin \pi T\tau/(\pi T)$ 
(cf.Ref.~\cite{Shankar90}) yielding 
\begin{eqnarray} 
F(\omega)\stackrel{T\ll\Omega}{\longrightarrow}\,\, &&
2\Re \left[ \frac{e^{-i\pi r}}{\tilde\Omega^{2r}}
\int_{0}^{\infty}dt
e^{i\omega t}
\left( \frac{\pi T}{\mbox{sinh}\, \pi T t} \right)^{2r} \right]  \nonumber \\
&&=\frac{e^{\omega/2T}}{\tilde\Omega} 
\frac{2\pi}{\Gamma(2r)} 
\left( \frac{2\pi T}{\tilde\Omega} \right)^{2r-1}\, 
\left| \Gamma\left( r+\frac{i\omega}{2\pi T} \right)\right|^2
\, . 
                                                      \label{A5}
\end{eqnarray}


\begin{references}


\bibitem{Altshuler79}B.~L. Altshuler, and A.~G. Aronov, 
Solid State Commun. {\bf 30}, 115 (1979); B.~L. Altshuler,  A.~G. Aronov, 
and P.~A.~Lee, Phys. Rev. Lett. {\bf 44}, 1288 (1980); 

\bibitem{Altshuler85}B.~L. Altshuler, and A.~G. Aronov.
In A.~J. Efros and M.~Pollak, editors, {\em Electron--Electron
Interaction In Disordered Systems}, pp. 1--153. Elsevier Science Pub. 
  B. V., North--Holland, 1985.
  
  
\bibitem{Shon90}G.~Sch\"on, and A.~D.~Zaikin, Phys. Rep. {\bf 198}, 237 (1990). 



\bibitem{Likharev89}K.~Mullen, Y. Gefen, and 
E. Ben--Jacob,  Physica {\bf B 152}, 172 (1988); 
D. V. Averin, and K. K. Likharev,  
in {\em Mesoscopic Phenomena in Solid} editors  
B.~L. Altshuler, P. A. Lee, and R. A. Webb. Elsevier Science Publishers
  B. V., North--Holland, 1991, pp. 173--271. 

\bibitem{Nazarov89}Yu. V. Nazarov, Zh. Eksp. \& Theor. Fiz., {\bf 96}, 975 
(1989); [Sov. Phys. JETP {\bf 68}, 561 (1989)]. 

\bibitem{Levitov95}L.~S.~Levitov, and A.~V.~Shytov, (to be published). 

\bibitem{Devoret90}M. H. Devoret, D. Esteve, H. Grabert, G.-L. Ingold, 
H. Pothier, and C. Urbina, Phys. Rev. Lett., {\bf 64}, 1824 (1990). 

\bibitem{Glazman90} S. M. Girvin, L. I. Glazman, M. Jonson, D. R. Penn, 
M. D. Stiles, Phys. Rev. Lett., {\bf 64}, 3183 (1990). 

\bibitem{foot1}
We acknowledge very useful comments by M. Devoret on this point. Ref. 
\cite{Devoret90} as well as his lecture notes (Les--Houches, 1994, to be 
published) contain much of the physics alluded to above. 


\bibitem{Matveev91}K.~A.~Matveev, Zh. Eksp. \& Theor. Fiz., {\bf 99}, 1598 
(1991); [Sov. Phys. JETP {\bf 72}, 892 (1991)]. 

\bibitem{Averin92}D.~V.~Averin, and Yu. V. Nazarov, 
Phys. Rev. Lett., {\bf 65}, 2446 (1990). 

\bibitem{foot2}
This contribution is dominant only when the screening radius 
is much larger than the electron wavelength. 

\bibitem{foot3}
We shall not treat here the periodic structure as a function of $N_0$.
 
\bibitem{Negele}J. W. Negele, and H. Orland, {\em Quantum Many--Particle 
Systems}, Addison--Wesley publishing company, 1988. 

\bibitem{Fin}A.~M. Finkelstein, 
{\em Electron Liquid in Disordered Conductors}, volume~14 of {\em
  Soviet scientific reviews}, editor I. M. Khalatnikov, 
Harwood Academic Publishers GmbH, 1990; A.~M. Finkelstein, 
Physica {\bf B 197}, 636 (1994).


\bibitem{Abrikosov64}A. A. Abrikosov, L. P. Gorkov, and I. E. Dzyaloshinski, 
{\em Methods of Quantum Field Theory in Statistical Physics.} Prentice--Hall, 
New Jersey 1963. 

\bibitem{foot4}
We note the connection to the dynamically screened interaction, 
$V({\bf Q},\omega_m)$, discussed in Section \ref{s2}, cf. Eq.~(\ref{scr}). 
Making the extension $D{\bf Q}^2\rightarrow D{\bf Q}^2+\gamma$ ($\gamma$ 
being a phenomenological leakage rate of the global charge in the dot) and 
taking ${\bf Q}=0$ we obtain 
$$
\left[ V({\bf Q}=0, \omega_m) \right]^{-1}=
\left[ V^{[0]} \right]^{-1}+ 
\left[ \Delta + e^2 R |\omega_m| \right]^{-1}, 
$$
where $V^{[0]}=e^2/C$ and $e^2 R\equiv \Delta/\gamma$. This coincides with 
Eq.~(\ref{Vstr}) for $\Delta\ll e^2/C$.   

\bibitem{Spivak94} B. Spivak, unpublished (1990). 

\bibitem{foot5}
Although Eq. (\ref{sas}) is supposed to be valid only for 
$\Omega\tau\gg 1$, it has an almost correct behavior for small $\tau$ as 
well. As one expects, the short time ($\Omega\tau\ll 1$) behavior of the 
action corresponds to an  unscreened interaction, cf. Eq.~(\ref{S0}), 
$S(\tau)\approx 0.5 V^{[0]}|\tau|$, 
whereas Eq. (\ref{sas}) leads to $S(\tau)\approx 0.57 V^{[0]}|\tau|$. 
Consequently our results, based on  Eq. (\ref{sas}) are expected to be 
qualitatively correct even for $\epsilon\gg \Omega$.

\bibitem{Shankar90}R.~Shankar, Int. J. Mod. Phys. {\bf B 4}, 2371 (1990); 
C.~de C.~Chamon, D.~E.~Freed, and X.~G.~Wen, Phys. Rev. {\bf B 51}, 2363 
(1995). 

\end{references}
\end{document}